\documentclass[aps,prl,twocolumn,amsmath,amssymbs,superscriptaddress]{revtex4-1}

\usepackage{graphicx}
\usepackage{amsmath,amsfonts,amssymb}
\usepackage{epsfig}
\usepackage{wrapfig}
\usepackage[usenames]{color}
\usepackage{array}
\usepackage{times}
\usepackage{float}
\usepackage{epstopdf}

\def\wo{\omega_{0}}

\def\dv{\delta\nu}

\def\a{a}
\def\ad{a^{\dag}}

\def\b{b}
\def\bd{b^{\dag}}

\begin{document}

\title{Photonic Anomalous Quantum Hall Effect}

\author{Sunil Mittal}
\email{mittals@umd.edu}
\affiliation{Joint Quantum Institute, NIST/University of Maryland, College Park, MD 20742, USA}
\affiliation{Department of Electrical and Computer Engineering, and IREAP, University of Maryland, College Park, MD 20742, USA}

\author{Venkata Vikram Orre}
\affiliation{Joint Quantum Institute, NIST/University of Maryland, College Park, MD 20742, USA}
\affiliation{Department of Electrical and Computer Engineering, and IREAP, University of Maryland, College Park, MD 20742, USA}

\author{Daniel Leykam}
\affiliation{Center for Theoretical Physics of Complex Systems, Institute for Basic Science (IBS), Daejeon 34126, Republic of Korea}

\author{Y. D. Chong}
\affiliation{Division of Physics and Applied Physics, School of Physical and Mathematical Sciences, Nanyang Technological University, Singapore 637371, Singapore}
\affiliation{Centre for Disruptive Photonic Technologies, Nanyang Technological University, Singapore 637371, Singapore}

\author{Mohammad Hafezi}
\affiliation{Joint Quantum Institute, NIST/University of Maryland, College Park, MD 20742, USA}
\affiliation{Department of Electrical and Computer Engineering, and IREAP, University of Maryland, College Park, MD 20742, USA}
\affiliation{Department of Physics, University of Maryland, College Park, MD 20742, USA}

\begin{abstract}
We experimentally realize a photonic analogue of the anomalous quantum Hall insulator using a two-dimensional (2D) array of coupled ring resonators. Similar to the Haldane model, our 2D array is translation invariant, has zero net gauge flux threading the lattice, and exploits next-nearest neighbor couplings to achieve a topologically non-trivial bandgap. Using direct imaging and on-chip transmission measurements, we show that the bandgap hosts topologically robust edge states. We demonstrate a topological phase transition to a conventional insulator by frequency detuning the ring resonators and thereby breaking the inversion symmetry of the lattice. Furthermore, the clockwise or the counter-clockwise circulation of photons in the ring resonators constitutes a pseudospin degree of freedom. We show that the two pseudospins acquire opposite hopping phases and their respective edge states propagate in opposite directions. These results are promising for the development of robust reconfigurable integrated nanophotonic devices for applications in classical and quantum information processing.
\end{abstract}

\maketitle


Photonics has emerged as a versatile platform to explore model systems with nontrivial band topology, a phenomenon originally associated with condensed matter systems \cite{Lu2014, Ozawa2019}. For example, photonic systems have realized analogues of the integer quantum Hall effect \cite{Raghu2008, Wang2009, Hafezi2011, Hafezi2013, Mittal2014}, Floquet topological insulators \cite{Rechtsman2013, Minkov2016, Maczewsky2017, Mukherjee2017},  quantum spin-Hall and valley-Hall phases \cite{Khanikaev2013, Cheng2016, Ma2016, Gao2017, Noh2018}, as well as topological crystalline insulators \cite{Wu2015, Barik2018, Shalaev2019}. From an application perspective, the inherent robustness of the topological systems has enabled the realization of photonic devices that are protected against disorder, such as optical delay lines \cite{Hafezi2013, Mittal2014}, lasers \cite{St-Jean2017, Bahari2017, Bandres2018}, quantum light sources \cite{Mittal2018}, and quantum-optic interfaces for light-matter interactions \cite{Barik2018}. At the same time, features unique to bosonic systems, such as the possibility of introducing gain and loss into the system \cite{Rudner2009, Zeuner2015, Leykam2017, Yao2018, Gong2018}, parametric driving, and squeezing of light \cite{Peano2016, Peano2016, Shi2017, Mittal2018}, have provided an opportunity to explore topological phases that cannot be realized in fermionic systems.

Despite these advances, there has not yet been a nanophotonic realization of the anomalous quantum Hall phase -- a two-dimensional Chern insulator with zero net gauge flux \cite{Haldane1988, Kane2005}. This is noteworthy because the various topological phases differ significantly in the origin of non-trivial band topology, and therefore offer different forms of topological protection. For instance, topological edge states in valley-Hall and topological crystalline insulator lattices manifest on internal boundaries between ``opposite'' domains instead of external edges \cite{Ma2016, Wu2015}, and are protected only against certain boundary deformations (e.g., 120$^\circ$ bends but not 90$^\circ$ bends) \cite{Ma2016, Wu2015}. The quantum Hall and anomalous quantum Hall phases, by contrast, are significantly more robust: topological edge states can appear along external edges, and are protected irrespective of the lattice shape.  Moreover, whereas the quantum Hall phase requires a non-zero net gauge flux \cite{Hafezi2011, Hafezi2013, Mittal2014}, the anomalous quantum Hall phase can occur in a periodic lattice with zero net flux. The translational invariance of anomalous Hall lattices is an advantage, because a topological-to-trivial phase transition can be induced simply by introducing on-site potentials \cite{Haldane1988}.

\begin{figure*}
\centering
\includegraphics[width=0.98\textwidth]{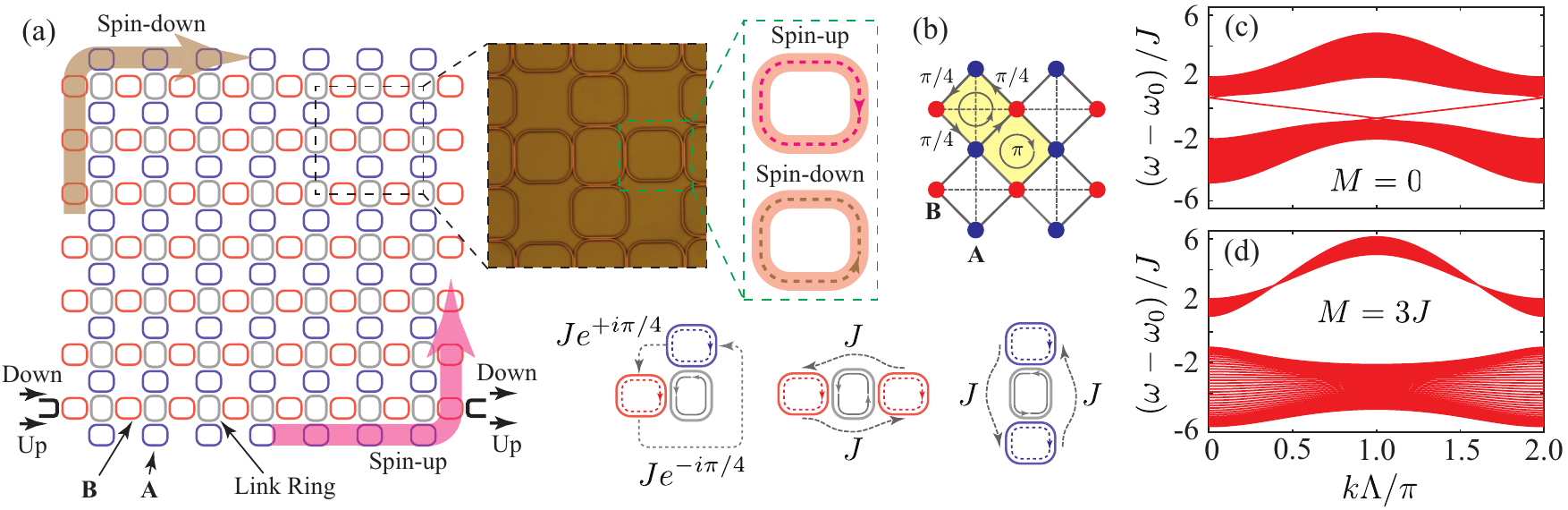}
\caption{
(a) Schematic of the 2D array of ring resonators, with site-rings \textbf{A} and \textbf{B} (shown in blue and red, respectively) coupled using link-rings (grey). The input and the output waveguides are shown in black. Top-left inset: microscope image of the device. Top-right inset: the ring resonators support a pseudospin degree of freedom, up and down, which correspond to the clockwise and the counterclockwise circulation of photons in the site rings, respectively. The choice of the input and the output ports allows us to selectively excite and measure the pseudospin up/down and the corresponding topological edge states travel in opposite directions (pink and brown arrows, respectively). Bottom inset: schematics for nearest-neighbor hopping (left) and next-nearest-neighbor hoppings (center and right) for the pseudospin-up. (b) Schematic of the 2D lattice describing the resonator array. Red and blue circles indicate \textbf{A} and \textbf{B} lattice sites respectively. Solid lines denote nearest-neighbor hoppings between \textbf{A} and \textbf{B} sites, with hopping phases indicated. Dashes indicate next-nearest neighbor hoppings. The gauge flux is $\pm \pi$ in a single plaquette, and zero over a unit cell of 2 plaquettes (shaded yellow). (c)--(d) Band diagram of the semi-infinite lattice for $M=0$ and $M = 3J$, respectively, where $J$ is the coupling strength between the lattice sites and $k\Lambda$ is the phase between neighboring site rings. For $M<2J$, the lattice is topological and exhibits edge states. The lattice is topologically trivial when $M>2J$.
}
\label{fig:1}
\end{figure*}

In this work, we demonstrate a nanophotonic analogue of the anomalous quantum Hall system using a periodic 2D checkerboard lattice of coupled ring resonators with nearest and next-nearest neighbor couplings. As proposed in Ref.~\cite{Leykam2018}, the tight-binding description of the photonic lattice is similar to the Haldane model \cite{Haldane1988}, in that the net gauge flux threading the lattice is zero, but next-nearest neighbor couplings induce non-zero local gauge flux. This effectively breaks time reversal symmetry and creates a topologically nontrivial band gap. We directly image the light intensity distribution in the lattice, revealing topological edge states in this gap that are robust against missing-site defects and propagate around 90$^\circ$ corners without any scattering into the bulk. Because the overall structure is time-reversal invariant, it hosts a pseudospin degree of freedom associated with the clockwise and the counter-clockwise (time-reversed) propagation of photons in the rings. By selective excitation of the pseudospins, we show that time-reversal invariance is effectively broken within each decoupled pseudospin sector, similar to the Kane-Mele quantum spin Hall model with no Rashba coupling \cite{Kane2005}, and the edge states associated with the two pseudospins propagate in opposite directions. Furthermore, we demonstrate a transition between topologically nontrivial and trivial phases by simply detuning the ring resonance frequencies, and observe edge states at an internal boundary between the two phases. We note that the system is periodic and does not require staggering the phases of the couplings, unlike the coupled-resonator system of Refs.~\cite{Hafezi2011, Hafezi2013, Bandres2018} which realizes the integer quantum Hall effect. These features are highly promising for the development of topological nanophotonic devices that can be dynamically reconfigured via optical, electrical, or thermal pumping \cite{Shalaev2018, Leykam2018}. Reconfigurability of topological systems has been demonstrated at microwave frequencies \cite{Cheng2016}, but is yet to be achieved in the optical regime.


Our system, shown in Fig.~\ref{fig:1}(a), consists of two interposed square lattices of ring resonators, with respective sites labelled \textbf{A} and \textbf{B}~\cite{Leykam2018}. These site-ring resonators are coupled to their neighbors and also next-nearest neighbors using another set of rings, the link rings. The resonance frequencies of the link rings are detuned from those of the site-rings by one half free-spectral range by introducing an extra path-length such that the round-trip phase at site-ring frequencies is $\pi$ \cite{Hafezi2011, Hafezi2013}. Therefore, the link rings introduce a direction-dependent hopping phase $\pm \pi/4$ when the photons hop from one lattice site to their nearest neighbors, while the hopping phase for next-nearest neighbors is zero (Fig.\ref{fig:1}(a)). As a result, the local effective magnetic flux (gauge flux) threading a plaquette (of two \textbf{A} and two \textbf{B} site rings) is $\pm \pi$, whereas the net flux threading a unit cell (of two plaquettes) is zero (Fig.\ref{fig:1}(b)). This staggered flux arrangement, originally conceived by Haldane, effectively breaks time-reversal symmetry and gives rise to an anomalous quantum Hall phase without Landau levels \cite{Haldane1988}.

\begin{figure*}
\centering
\includegraphics[width=0.98\textwidth]{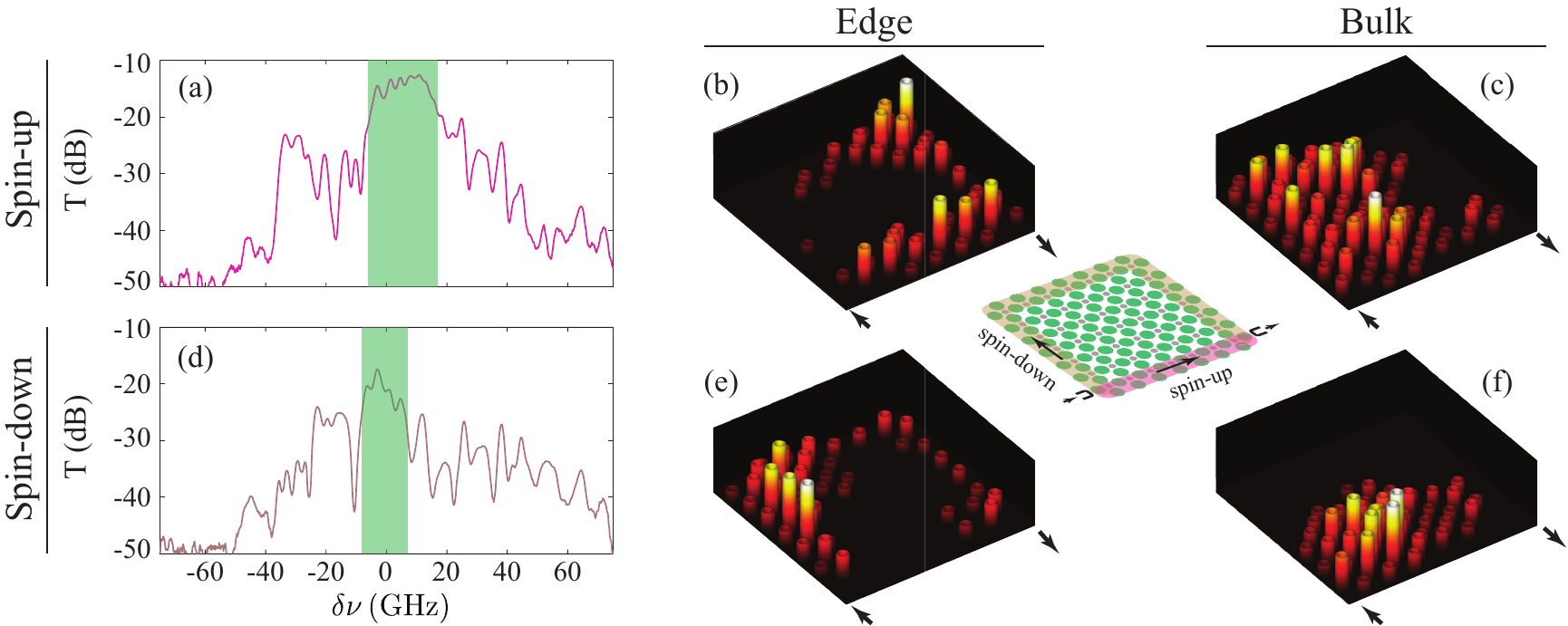}
\caption{
(a) Measured transmission (T) spectrum for the topologically nontrivial lattice ($M = 0$) with the pseudospin-up excitation. The green shaded region indicates the frequency band over which topological edge states are observed in direct imaging.  (b) The corresponding spatial intensity distribution obtained through direct imaging at $\dv \approx 0$ (integrated over a frequency range of $5~\text{GHz}$). Edge states are found to circulate CCW around the lattice. (c) Intensity distribution at $\dv \approx -20$ GHz, showing scattering into the bulk. (d)--(f) The corresponding results for the pseudospin-down excitation. The edge state now travels CW, and the measured transmission is about 5 dB lower due to the longer path from input to output port. All spatial intensity distributions show only site ring resonators}
\label{fig:2}
\end{figure*}

The photonic lattice is time-reversal invariant, and supports a pseudospin (up or down) degree of freedom associated with the circulation direction (clockwise or counter-clockwise) of photons in the site-ring resonators (Fig.\ref{fig:1}(a)). The two pseudospins are time-reversed partners, and thus identical in terms of their coupling constants and resonance frequencies. Therefore, they realize two copies of the anomalous Hall phase \cite{Kane2005}. However, they differ in the hopping phase between nearest-neighbors, which indicates that time-reversal symmetry is effectively broken for each pseudospin. Specifically, if the hopping phase for the  pseudospin-up is $+\pi/4$, the corresponding phase for the pseudospin-down is $-\pi/4$. The tight-binding Hamiltonian describing the system is \cite{Leykam2018}
\begin{eqnarray}
H &=& \sum_{i,j,\sigma} \left(\wo - M\right) \ad_{i,\sigma} \a_{i,\sigma} + \left(\wo + M\right) \bd_{i,\sigma} \b_{i,\sigma} \\
  &-& J \left( \ad_{j,\sigma} \a_{i,\sigma} + \bd_{j,\sigma} \b_{i,\sigma} + \ad_{j,\sigma} \b_{i,\sigma} e^{-i \sigma \phi_{i,j}}  + \mathrm{h.c.} \right). \nonumber
\end{eqnarray}
Here, $a_{i,\sigma}, b_{i,\sigma}$ are the annihilation operators corresponding to site rings \textbf{A} and \textbf{B}, respectively, at lattice site index $i = \left(x,y\right)$. $\sigma = \pm 1$ is the pseudospin index for the up/down spins, respectively. $J$ is the coupling strength between nearest and the next-nearest neighbor sites, and $\phi_{i,j} = \pm \pi/4$ is the direction-dependent hopping phase between sites \textbf{A} and \textbf{B}, as shown in Fig.~\ref{fig:1}(a). We include a frequency detuning $M$ between the \textbf{A} and \textbf{B} site rings. When $M<2J$, the lattice band structure hosts a topological band gap, occupied by unidirectional, topologically-robust edge states (Fig.~\ref{fig:1}(c)). By contrast, when $M>2J$, the lattice is topologically trivial and the edge states are absent (Fig.~\ref{fig:1}(d)). Furthermore, because of the spin-dependent hopping phase, the edge states corresponding to the two spins propagate around the lattice in opposite directions. With the two pseudospins decoupled, this is similar to the Kane-Mele model of the quantum spin Hall effect \cite{Kane2005}. We note that in our system the nearest and next-nearest neighbor couplings have the same strength $J$. Because the width of the topological band gap is dictated by the next-nearest neighbor couplings, the topological band width in our system is of the order of the coupling strength, unlike the Floquet systems where the width of the topological band is limited by the frequency of modulation \cite{Rechtsman2013}.

We used the silicon-on-insulator platform to realize our topological system \cite{Hafezi2013, Mittal2014, Mittal2016}. This platform is compatible with the existing CMOS fabrication process and works at telecom wavelengths. The ring resonator waveguides are 510 nm wide and 220 nm high, and support a single transverse-electric (TE) mode. The coupling gap between resonators is 180 nm, with coupling strength $J$ estimated to be $15.6(4)\, \text{GHz}$. To probe the lattice we couple input and output waveguides as shown in Fig.~\ref{fig:1}(b). Specifically, we couple a tunable, continuous-wave laser at the input port and measure the power transmission at the output port. We use a microscope objective to collect photons scattered by the ring resonator waveguides. This allows us to directly image the path followed by light in the lattice as we tune the input wavelength. Furthermore, by choosing the input and the output ports, we can selectively excite and measure a given pseudospin. The use of directional couplers and low back-scattering in the resonator waveguides ensures that mixing between the two pseudospins is negligible.

\begin{figure*}
  \centering
  \includegraphics[width=0.98\textwidth]{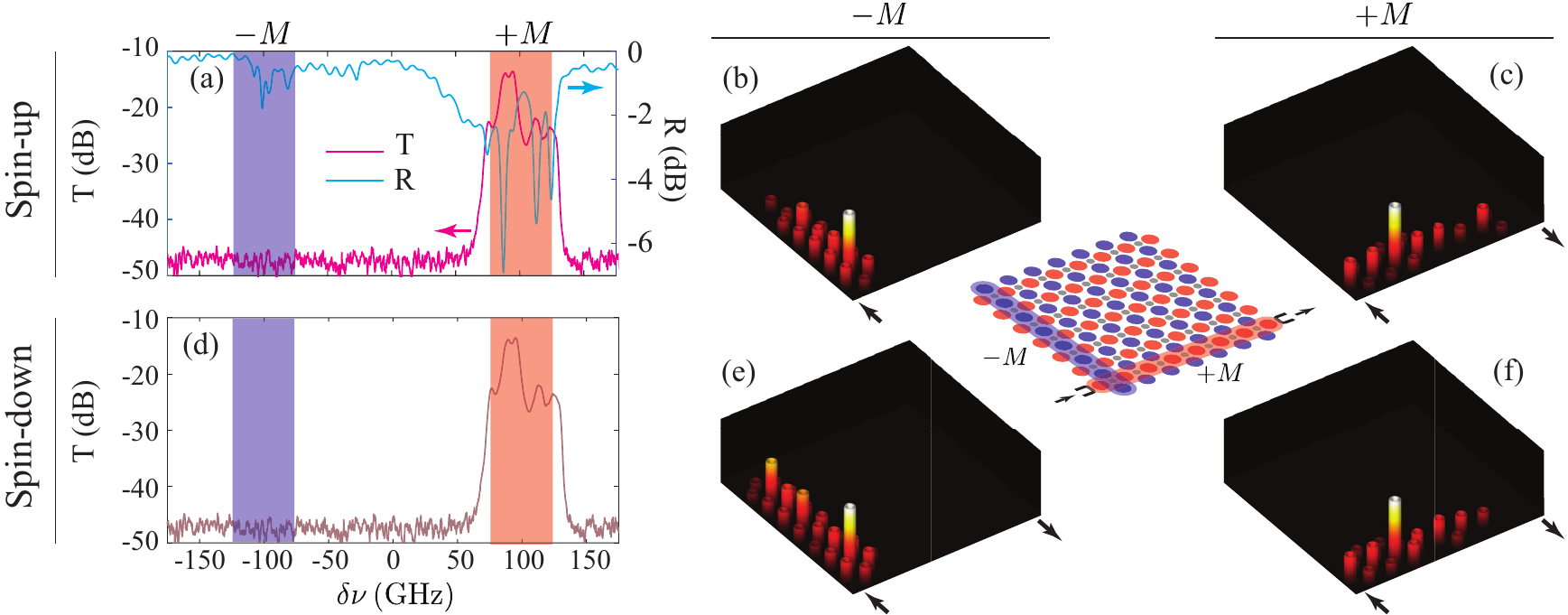}
  \caption{
    (a) Measured transmission (T) and reflection (R) spectrum of a topologically trivial device, with $M \approx 98\,\text{GHz} \gg 2 J$ and pseudospin-up excitation. The transmission is negligible within the bandgap ($\dv \approx 0$) due to the absence of edge states. The transmission peak (and reflection dip) at $\dv = +M$ coincides with the resonance frequency of the \textbf{B} rings.  At $\dv = -M$, the lattice absorbs some light from the input but the light does not reach the output because of the large frequency mismatch between \textbf{A} and \textbf{B} rings. (b,c) Spatial intensity profiles for $\dv = +M$ and $\dv = -M$, showing excitation of only \textbf{B} and \textbf{A} rings, respectively. (d) Measured transmission spectrum for the pseudospin-down excitation. Unlike the topological case, the transmission is same irrespective of the spin. (e,f) Spatial intensity distributions also remain almost identical, confirming topologically trivial nature of the lattice.
  }
  \label{fig:3}
\end{figure*}

To observe topological edge states, we fabricated an array of 56 \textbf{A} resonators and 56 \textbf{B} resonators, as shown schematically in Fig.~\ref{fig:1}(a). For this device we choose $M = 0$, that is, the \textbf{A} and \textbf{B} resonators are identical, corresponding to the non-trivial topological phase. Fig.~\ref{fig:2}(a) shows the measured transmission spectrum at the lattice output for the pseudospin-up excitation. We observe high transmission near the frequency detuning $\dv \approx 0$. Fig.~\ref{fig:2}(b) shows the measured spatial intensity profile at $\delta\nu \approx 0$, integrated over a frequency range of $5$ GHz. The light is confined to the lattice edge and propagates around the lattice in a counter-clockwise direction. Furthermore, the light travels around two sharp $90^{o}$ bends without scattering into the bulk of the lattice. This shows that this high-transmission region around $\dv \approx 0$ is indeed the topological edge band. The decrease in light intensity as it propagates along the edge is mainly due to scattering losses in the resonator waveguides. By contrast, when we excite the lattice outside this band, for example at $\delta\nu \approx -20$ GHz, the spatial intensity distribution occupies the bulk of the lattice, as shown in Fig.~\ref{fig:2}(c). Moreover, the spatial intensity profile in the bulk band is sensitive to even small changes in the excitation frequency whereas the intensity profile in the edge band is relatively constant throughout the edge band. Note that the circulation direction (CCW) around the lattice is opposite to the circulation direction (CW) in the site ring resonators.

This observation of topological edge states is also a demonstration of their robustness against fabrication-induced disorder. Although the fabrication was performed at a state-of-the-art commercial foundry (IMEC, Belgium), there is significant disorder in the ring resonance frequencies $\left(\Delta U \right)$, which we measured to be around $33\,\text{GHz}$, comparable to the band gap width of around $2J = 32 \text{GHz}$.

Next, we probe the spin-polarized nature of the topological edge states by exciting the lattice with the pseudospin-down. Fig.~\ref{fig:2}(d) shows the resulting transmission spectrum. The measured spatial intensity profile at $\dv \approx 0$ reveals an edge state which now propagates around the lattice in a clockwise direction (Fig.~\ref{fig:2}(e)). Again, the edge state intensity is confined to the physical edge of the lattice. The transmission at $\dv \approx 0$ is approximately 5 dB lower than in the pseudospin-up case, because the edge state for the pseudospin-down travels a much longer path between the input and output couplers. At frequencies outside the band gap, we again see scattering into the bulk (Fig.~\ref{fig:2}(f)).

To demonstrate the existence of a topological phase transition, we fabricated another device with sublattice detuning $M \approx 98$ GHz, significantly larger than the transition threshold of $2J \approx 31 \text{GHz}$. This detuning is achieved by increasing (decreasing) the length of the \textbf{A} (\textbf{B}) ring resonators by 30 nm, which red (blue) shifts their resonance frequencies and results in a non-zero value of $M$. Fig.~\ref{fig:3}(a) shows the measured transmission spectrum at the output, for the pseudospin-up input. We observe almost negligible transmission at $\dv = 0$, indicating the absence of any transmitting channels in the bandgap. There is a single transmission band at $\dv \approx M \approx 100\,\textrm{GHz}$. The measured spatial intensity distribution at $\dv = 100\,\textrm{GHz}$, shown in Fig.~\ref{fig:3}(c), reveals only a few \textbf{B} rings (which are resonant with the input frequency) are excited near the input port. In this regime, the \textbf{A} and \textbf{B} rings are very weakly coupled due to the large resonance frequency mismatch. The transmission is negligible at $\dv = -100 ~\text{GHz}$ (the resonance frequency of the \textbf{A} rings) because the input and output ports are coupled to \textbf{B} rings. A small amount of absorption by the \textbf{A} rings is visible in the reflection spectra shown in Fig.~\ref{fig:3}(a), and in the spatial intensity profile of Fig.~\ref{fig:3}(b). More importantly, we find that flipping the spin of excitation does not affect the transmission spectrum or the spatial intensity profile, as shown in Fig.~\ref{fig:3}(d-f); this confirms the lattice is topologically trivial.

\begin{figure}
\centering
\includegraphics[width=0.48\textwidth]{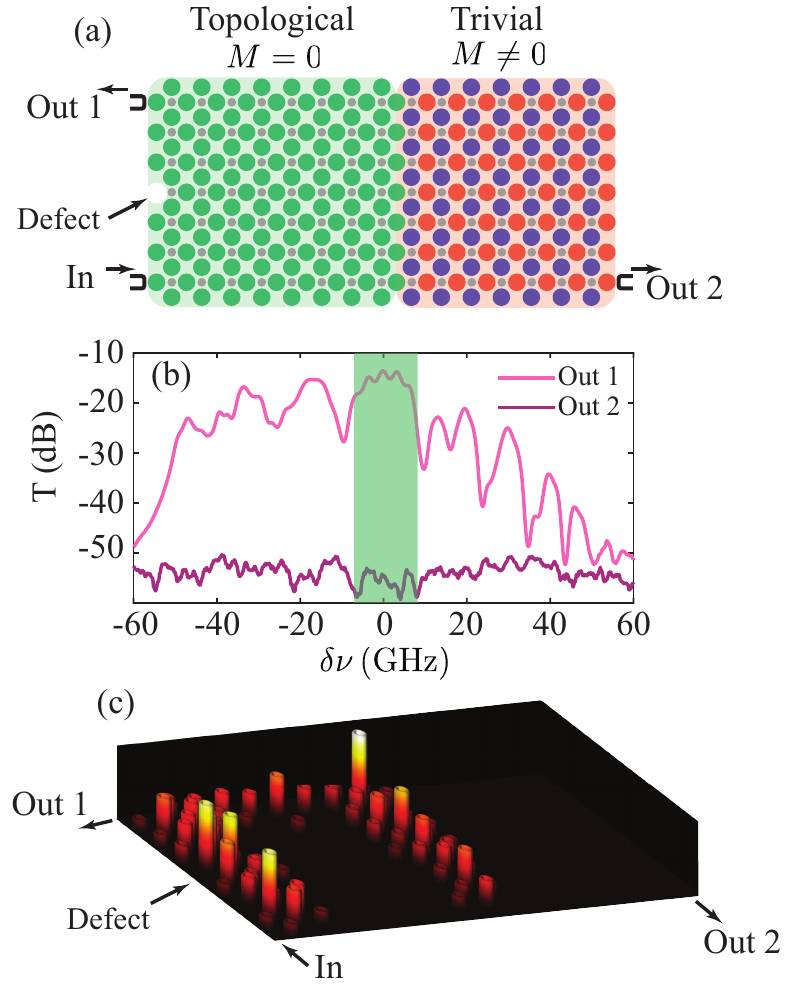}
\caption{(a) Schematic of device with an interface between topological $\left(M = 0 \right)$ and trivial $\left(M > 2J \right)$ domains. The topological domain also hosts a defect in the form of a missing site-ring resonator. (b) Measured transmission (T) spectrum from input to the two output ports, for the pseudospin-down excitation. The light follows the interface, leading to negligible output at the trivial domain coupler. (c) Measured spatial intensity
profile.}
\label{fig:4}
\end{figure}

To verify that the edge states are not artifacts of the physical lattice boundary, we fabricated a device with an interface between a topological lattice $\left( M = 0 \right)$ and a trivial lattice $\left(M \approx 98\, \text{GHz} \right)$, shown in Fig.~\ref{fig:4}(a). We place an input port on one edge of the topologically nontrivial domain and monitor two output ports on the edges of the nontrivial and trivial domains. The measured transmission spectra at the two output ports, with the pseudospin-down excitation, are shown in Fig.~\ref{fig:4}(b).  At frequencies within the bandgap of the nontrivial domain (highlighted in Fig.~\ref{fig:4}(b)), we observe edge states propagating clockwise around the nontrivial domain (Fig.~\ref{fig:4}(c)). These edge states then follow the ``internal'' domain boundary, and do not enter the topologically trivial domain; accordingly, negligible transmission is observed at the output port in the trivial domain.  As a further test of robustness, we deliberately removed one site-ring resonator from the edge of the topologically nontrivial domain, as indicated in Fig.~\ref{fig:4}(a) and (c). The edge state routes around the defect, without scattering into the bulk. We emphasize that this topological protection is superior to recently-demonstrated crystalline symmetry-protected and valley Hall topological edge states, which are sensitive to symmetry-breaking disorder~\cite{Barik2018, Gorlach2018, Noh2018, He2019, Shalaev2019}.


To summarize, we demonstrated topologically robust edge states in a nanophotonic analogue of the anomalous quantum Hall effect, using a periodic 2D lattice of ring resonators with zero net gauge flux. We showed a topological-to-trivial phase transition, induced by relatively small detunings of the ring resonance frequencies. In the future, this phase transition can be utilized for robust routing and switching of light in integrated photonic devices \cite{Leykam2018}. Specifically, the silicon photonics platform can easily include active components, such as metal heaters \cite{Mittal2016} or electro-optic modulators \cite{Reed2010} to dynamically tuning the ring resonances. Moreover, the large Kerr nonlinearity of silicon could be leveraged for robust, optically-reconfigurable light routing, and to explore the behavior of topological states in a nonlinear regime.

This research was supported by the AFOSR-MURI grant FA9550-16-1-0323, the Physics Frontier Center at the Joint Quantum Institute, the Institute for Basic Science in Korea (IBS-R024-Y1), the Singapore MOE Academic Research Fund Tier 2 Grant MOE2015-T2-2-008, and the Singapore MOE Academic Research Fund Tier 3 Grant MOE2016-T3-1-006.

\bibliographystyle{apsrev}
\bibliography{Anom_Hall_Biblio}

\end{document}